\documentclass[10pt,letterpaper]{article}

\usepackage{sectsty}
\usepackage{graphicx}
\usepackage[hypertexnames=false]{hyperref}
\usepackage[dvipsnames]{xcolor}
\usepackage{cite}
\usepackage{amsfonts}
\usepackage{amsmath}
\usepackage{stmaryrd}
\usepackage{enumitem}
\usepackage{bm}
\usepackage{amssymb}
\usepackage[ruled]{algorithm2e}
\usepackage{booktabs}
\usepackage[font={footnotesize}]{caption}
\allowdisplaybreaks

\newtheorem{definition}{Definition}	

\renewcommand{\O}{\mathcal{O}}

\newcommand{\V}{\mathcal{V}}
\newcommand{\E}{\mathcal{E}}
\newcommand{\G}{\mathcal{G}}

\newcommand{\gt}{$\mathcal{G}(\mathcal{V}, \mathcal{E}, \mathcal{W})$~}
\newcommand{\R}{\mathbb{R}}
\newcommand{\ie}{i.e.}

\newcommand{\ptv}{\texttt{EDRep~}}
\renewcommand{\L}{\mathcal{L}}
\renewcommand{\P}{\mathbb{P}}

\hypersetup{colorlinks,linkcolor={magenta},citecolor={blue},urlcolor={magenta}}

\title{An embedding-based distance for temporal graphs}

\author{Lorenzo Dall'Amico~$^{1,*}$, Alain Barrat~$^{2, \dagger}$, Ciro Cattuto~$^{1, \dagger}$\\\vspace{-8pt}
	\\{\small {$^{1}$}ISI Foundation, Turin, 10126, Italy}
	\\{\small{$^{2}$}Aix-Marseille Univ, Université de Toulon, CNRS, CPT, Marseille 13009, France}\\
	{\small{{$^{*}$}Corresponding to: \tt{lorenzo.dallamico@isi.it}}}\\
	{\small $^{\dagger}$Equal contribution}}

\date{\today}

\begin{document}

\maketitle

\begin{abstract}
	Temporal graphs are commonly used to represent time-resolved relations between entities in many natural and artificial systems. Many techniques were devised to investigate the evolution of temporal graphs by comparing their state at different time points. However, quantifying the similarity between temporal graphs as a whole is an open problem. Here, we use embeddings based on time-respecting random walks to introduce a new notion of distance between temporal graphs. This distance is well-defined for pairs of temporal graphs with different numbers of nodes and different time spans. We study the case of a matched pair of graphs, when a known relation exists between their nodes, and the case of unmatched graphs, when such a relation is unavailable and the graphs may be of different sizes. We use empirical and synthetic temporal network data to show that the distance we introduce discriminates graphs with different topological and temporal properties. We provide an efficient implementation of the distance computation suitable for large-scale temporal graphs.
\end{abstract}

\section*{Introduction}

Graphs are ubiquitous mathematical entities formed by a set of \emph{nodes} and one of \emph{edges} connecting pairs of nodes~\cite{barabasi1999emergence,newman2003structure, Barrat:2008}. They can model a wide range of interacting systems, such as social~\cite{scott1988trend, wasserman1994social}, technological~\cite{amaral2000classes}, spatial~\cite{barthelemy2011spatial} and biological networks~\cite{alm2003biological}, and owe their popularity to their ability to encode the complex relational structure of these systems. Real-world systems, moreover, often have temporal properties that cannot be encoded in static graphs and call for modeling based on time-resolved network representations, known as \emph{temporal graphs}~\cite{holme2012temporal}. Examples of these include transportation~\cite{zhao2019t} and ecological networks~\cite{blonder2012temporal}, human close-range interactions~\cite{eagle2009inferring, cattuto2010dynamics, barrat2013empirical, barrat2013temporal, stopczynski2014measuring}, collaboration networks~\cite{bravo2019gender, cazabet2018tracking, carstensen2015spatio}, etc.

\medskip
Given their ubiquitous use in representing such diverse kinds of systems, it is crucial to be able to compare them. In fact,
defining and computing a similarity measure between pairs of graphs~\cite{bellet2015metric} underpins many important applications and tasks, including machine learning tasks such as graph classification. However, given the high dimensionality of graphs and their structural complexity, many different notions of distance can be devised, capturing different properties of interest. Hence, many definitions of distance for static graphs were introduced, as reviewed in Refs.~\cite{wills2020metrics, tantardini2019comparing, hartle2020network, barros2021survey, ma2021deep}. One of the most straightforward approaches is the \emph{edit distance}~\cite{sanfeliu1983distance}, which counts the number of elementary changes (edge or node removal or addition) needed to transform one graph into another and only accounts for local information.
Approaches that probe the global graph structure are usually based on matrix inversion~\cite{koutra2013deltacon, monnig2018resistance} but generally have a high computational cost and approximations are needed to achieve satisfactory computational performances. The above methods are all designed for comparing pairs of graphs with a known correspondence between nodes, a case we will refer to as \emph{matched} graphs. The more general case of \emph{unmatched} graphs entails defining distances for pairs of graphs for which a mapping between nodes is unavailable, and the graphs may have different numbers of nodes. In this case, one can either extract and compare a suitable set of chosen graph features, as done in Refs.~\cite{berlingerio2012netsimile, bagrow2019information}, or leverage the spectral properties of appropriate matrix representations~\cite{apolloni2011introduction, shimada2016graph, torres2018graph, tsitsulin2018netlsd}.  In this case, the computational complexity is also a bottleneck, and approximation heuristics are often required~\cite{wills2020metrics}.

\medskip

In the case of temporal graphs, the temporal dimension adds to the complexity of graph comparison. Many of the distance measures mentioned above for static graphs were used to study temporal graphs~\cite{donnat2018tracking}, typically to compare the state of the graph at different points in time and to identify anomalies and change points in the temporal evolution of the graph~\cite{masuda2019detecting,gelardi2019detecting,pedreschi2020dynamic, beladev2020tdgraphembed, huang2022learning}. Here instead, we study the problem of comparing temporal graphs, each taken as a whole. Unlike the snapshot-based methods mentioned above~\cite{masuda2019detecting,gelardi2019detecting, pedreschi2020dynamic, beladev2020tdgraphembed, huang2022learning}, we rely on a temporal embedding with a dimensionality that is independent of the temporal span of the graphs, so that we can compare temporal graphs with arbitrary, and different, temporal spans.
Few notions of distance were proposed to compare temporal graphs with arbitrary time spans. In particular, 
the method of Ref.~\cite{bail2023flow} is based on a set of features that can be tailored to a specific research question (similarly to \cite{berlingerio2012netsimile} for static graphs). 
Ref.~\cite{froese2020comparing} on the other hand addresses the same problem of temporal graph comparison we tackle here, but it assumes the graphs to be \emph{matched} and it is computationally inefficient (the graph dissimilarity is obtained solving an NP-hard optimization problem and the authors propose an approximation running in $\O(n^3)$ operations, where $n$ is the number of graph nodes).
Finally, Ref.~\cite{zhan2021measuring} only considers the \emph{unmatched} case and 
focuses on a specific property of temporal graphs, namely the comparison of temporal paths' statistics, to provide a definition of \emph{dissimilarity}.

\medskip

Here, we propose a computationally efficient method to compute an actual distance metric between pairs of temporal graphs, considering both the matched and unmatched cases. We build our comparison on top of graph embeddings, leveraging ideas from the local and global approaches mentioned above and encoding differences at the topological and temporal levels.
Commented \texttt{Python} code that implements the approach described here is publicly available at the link provided in Ref.~\cite{dallamico2024GDynaDist}.

\section*{Results}

Before delving into the details of our main contribution, we first state some basic definitions.

\begin{definition}[Temporal graph]
	\label{def:dynamic}
	A temporal graph \gt is a tuple $(\V, \E, \mathcal{W})$, where $\V$ is a set of $n$ nodes, $\E$ a set of temporal edges between pairs of nodes, and $\mathcal{W}$ a set of edge weights. For a discrete variable $t$ denoting time, all temporal edges $e \in \E$ have the form $e = (i, j, t)$, representing an interaction between nodes $i$ and $j$ at time $t$ and $w_e$ indicates the positive weight of edge $e$. We say an edge $(ij)$ is \emph{active} at time $t$ if $(i,j,t)\in \E$. A node $i$ is active at time $t$ if it belongs to an active edge at $t$.
\end{definition}

The graph \gt can be represented with a sequence of \emph{weighted adjacency matrices}, $\{W_t\}_{t =1,\dots,T}$, where $T$ is the number of time points. Each of these matrices has size $n \times n$ and entries $(W_t)_{ij} = w_{(i,j,t)} > 0$ if $(i,j,t) \in \E$ and $(W_t)_{ij} = 0$ otherwise. In the following, we will refer to these matrices as ``snapshots'' of the temporal graph at given times and to $T$ as the number of snapshots~\cite{rossetti2018community}. In a more general description of temporal networks, each interaction may be represented with a tuple $(i, j, t, \tau)$ where $\tau \in \R^+$ is an \emph{interaction duration}. We adopt the former notation because it is simpler to handle in the following. The two notations are completely interchangeable whenever the time is discrete and a finite temporal resolution $t_{\rm res}$ is set. In fact, all contacts with a duration within $t_{\rm res}$ are considered to be instantaneous, and $(i, j, t, \tau)$ can be represented by a set of interactions of unitary duration: $\{(i, j, t + \delta)\}_{\delta \in \{0,\dots,\tau-1\}}$. For this reason, Definition~\ref{def:dynamic} is sufficiently general, as it can also describe interactions lasting more than one temporal unit. 
For increasing values of $t_{\rm res}$, the temporal graph gets progressively more aggregated and loses information on the temporal ordering of contacts within $t_{\rm res}$, but partial information can be retained with edge weights expressing the fraction of $t_{\rm res}$ in which the edge was active.

\medskip

Let us now introduce the concept of \emph{matched} and \emph{unmatched} graphs.

\begin{definition}[Matched graphs]
	\label{def:matched}
	Let $\G_1(\V_1, \E_1,\mathcal{W}_1)$ and $\G_2(\V_2, \E_2, \mathcal{W}_2)$ be two temporal graphs with $n_1$ and $n_2$ nodes, respectively. We say that $\G_1$ and $\G_2$ are \emph{matched} if $n_1 = n_2$ and there exists a \emph{known} bijective function $\pi: \V_1 \to \V_2$ that maps each node of $\G_1$ to a node of $\G_2$. The two graphs are otherwise said to be \emph{unmatched}.
\end{definition}

When two graphs have the same size, \ie, the same number of nodes, there are always many possible mappings $\pi$ between their nodes, but for the \emph{matched} approach to be valuable, a known mapping associated with an external notion of node identity across the two graphs is necessary (e.g., nodes corresponding to the same persons in two temporal social networks). An implicit consequence of Definition~\ref{def:matched} is that all graph pairs with a different number of nodes are considered to be \emph{unmatched}. We now state the requirements for our distance measure and illustrate our main result.

\begin{figure*}[!t]
	\centering
	\includegraphics[width=\linewidth]{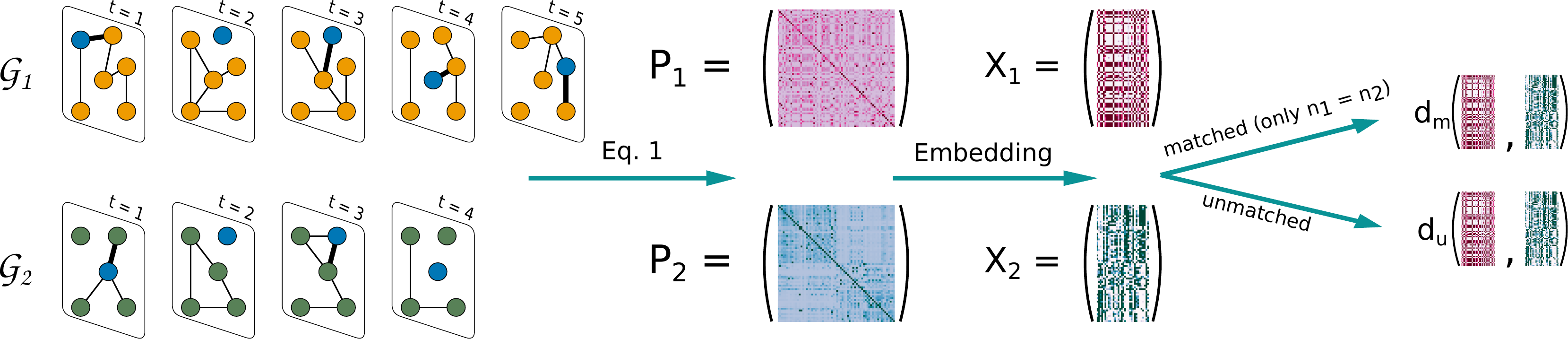}
	\caption{\textbf{Distance computation between two temporal graphs}. The inputs to the distance function are two temporal graphs $\G_1$ (top, orange nodes) and $\G_2$ (bottom, green nodes), generally with a different number of snapshots ($T_1 = 5$, $T_2 = 4$, in the example) and a different number of nodes $n$ (here $n_1 = 6$, $n_2 = 5$). Each graph is represented as a matrix $P \in\R^{n\times n}$, according to Eq.~\eqref{eq:Pdyn}, with entry $(ij)$ encoding the probability that a random walker goes from $i$ to $j$ following a time-respecting path (the walker's position at each time point is indicated in blue). The matrices $P_1, P_2$ are then embedded using the \texttt{EDRep} algorithm, mapping them to $X_1 \in \R^{n_1\times d}$ and $X_2 = \R^{n_2\times d}$. Finally, the matched -- Eq.~\eqref{eq:dm} -- and unmatched -- Eq.~\eqref{eq:du} -- distances are calculated. Notice that a necessary condition to compute $d_{\rm m}$ is that $n_1 = n_2$.}
	\label{fig:pipeline}
\end{figure*}

\medskip

In the following, we define two types of distance functions, $d_{\rm m}$, $d_{\rm u}$ between matched and unmatched graphs, respectively. The inputs to these functions are two temporal graphs, $\G_1$ and $\G_2$, with potentially different numbers of time points (snapshots) $T_1$ and $T_2$. In both cases, the distances should fulfill the following standard requirements~\cite{gromov1999metric}: i) non-negativity; ii) separation axiom; iii) symmetry; iv) triangle inequality. In addition, both distances should be able by design to detect differences induced by the permutation of time indices. The matched distance should also differentiate between temporal networks differing only by a permutation of node indices, while the unmatched one should be invariant under such transformation.

\medskip

Our proposed distances are computed in two steps:
i) we generate a temporal graph embedding, given by a matrix of size $(n \times d)$ (with $d$ independent of $T$), that encodes relevant topological and temporal properties {by leveraging time-respecting random walks on the temporal network};
ii) we define the distance in the embedded space, treating separately the matched and the unmatched cases.
The distance computation pipeline is illustrated in Fig.~\ref{fig:pipeline}. The steps outlined above are described in detail in the following two Sections titled~\nameref{sec:main.method} and \nameref{sec:main.dist}.

\subsection*{Temporal graph embedding}
\label{sec:main.method}
Graphs are rich data representations and it is challenging to define simple mathematical operations to manipulate them. Graph embeddings provide an interesting avenue for doing so, as they yield a representation in vector space that preserves some of the graphs' relevant properties~\cite{cai2018comprehensive, goyal2018graph, xu2021understanding, makarov2021survey}.
In the case of temporal graphs, the additional challenge of dealing with their temporal dimension has been tackled in many different ways: Ref.~\cite{gauvin2014detecting} represents time as an additional dimension of an adjacency tensor; in evolutionary spectral clustering~\cite{chi2007evolutionary, qin2016multi, liu2018global, xu2014adaptive, dall2020community}, a different embedding is obtained for each snapshot, and smoothness conditions are implemented on their temporal evolution; more recent work~\cite{zuo2018embedding, nguyen2018continuous, sato2019dyane} introduced approaches based on time-respecting random walks, which encode temporal properties relevant for dynamical processes occurring on temporal graphs; Refs.~\cite{beladev2020tdgraphembed, huang2022learning} also build one embedding per snapshot for studying the graph evolution and comparing different time points; the authors argue that a graph comparison metric should depend on the whole graph structure and thus define graph instead of node embeddings.

\medskip

Following an approach similar to Ref.~\cite{sato2019dyane}, we start from time-respecting random walks to generate temporal graph embeddings, which we use to define a distance based on global temporal graph properties. The embeddings are obtained as the solution to an optimization problem. We define a loss function whose argument is a transition matrix $P$ that is the limiting distribution of time-respecting random walks over the temporal graph
$\G$ to be embedded. $T$ being the number of snapshots of $\G$, 
we consider an ensemble of random walks of length $\ell$ drawn uniformly at random between $1$ and $T$, and starting from a randomly chosen node at time $t=T-\ell+1$;
at each time step, the walker, situated, e.g., on node $i$, can either move to a random neighbor of $i$ in the snapshot at time $t$ or stay in place on $i$; if the walker is situated on a node that is not active at time $t$, \ie, which has no neighbors in this snapshot, the walker stays in place with probability one. The graph snapshots on the left of Fig.~\ref{fig:pipeline} show examples of such time-respecting random walks on the input graphs.

\medskip

Specifically, given a temporal graph $\G$, with $W_t$ the snapshot adjacency matrices at times $t=1,\cdots,T$, we define for each snapshot $\hat{L}_t = (D_t + I_n)^{-1}(W_t + I_n)$, where $D_t$ is the degree matrix $D_t = {\rm diag}(W_t\mathbf{1}_n)$ of the snapshot, $I_n$ is the identity matrix of size $n \times n$, and $\mathbf{1}_n$ is the $n$-sized unity vector. We then define the matrix $P$ as
\begin{align}
        \label{eq:Pdyn}
 P = \frac{1}{T}\sum_{\tau = 1}^T \hat{L}_{\tau}\hat{L}_{\tau +1}\cdots \hat{L}_T \,\, ,
\end{align}
The matrices $\hat{L}_t$ and $P$ are sometimes referred to as \emph{temporal transition matrix} and \emph{global transition matrix}, respectively.
Time-respecting random walks depend on several properties of the temporal graph, that may include, for instance, the presence of a non-Markovian behavior, and of a broad distribution of the edge activity  and inter-event durations \cite{saramaki2015exploring}.
The limiting probability matrix $P$ of time respecting random walks encodes the aforementioned properties and it is sensitive to the permutation of time indices. We remark that the random walks we use can be of arbitrary length. Hence, the matrix $P$ is, in principle, sensitive to all time scales of the temporal graph at hand.

\medskip

Given the above probability matrix $P$ and a chosen embedding dimensionality $d$, we now generate, for each node $i$ of $\G$, a unitary node embedding vector $\mathbf{x}_i \in \R^d$.
To this end, we define $X \in \R^{n\times d}$ as a matrix where each row represents an embedding vector. We then express the loss function as:
    \begin{align}
        \label{eq:EDRep}
 \mathcal{L}_P(X) &= - \sum_{i, j \in \V}\left[P_{ij}~{\rm log}~Q_{ij}(X) - \frac{\mathbf{x}_i^T\mathbf{x}_j}{n}\right]\,\, , \\
 Q_{ij}(X) &= \frac{e^{\mathbf{x}_i^T\mathbf{x}_j}}{\sum_{k\in\V}e^{\mathbf{x}_i^T\mathbf{x}_k}} := \frac{e^{\mathbf{x}_i^T\mathbf{x}_j}}{Z_i}\,\, .
    \end{align}
 This loss function is the sum of the cross entropies between the distributions in the rows of $P$ and the variational distributions in the rows of $Q(X)$, with the addition of a regularization term. Since $Z_i$ is computed in $\O(n)$ operations for each $i$, the function $\mathcal{L}_P$ requires $\O(n^2)$ operations to be computed. We adopt the recently proposed \ptv algorithm~\cite{dall2023efficient} that introduces an efficient approximation of $Z_i$, optimizing $\L_P$ in $O(n)$ operations. For further details, we refer the reader to the \nameref{sec:supp} section.

\medskip

The matrix $X$ obtained by optimizing $\L_P$ defines the embedding of the temporal graph $\G$. In the following, we show how to leverage this embedding to define a distance between temporal graphs.

\subsection*{Defining an embedding-based distance}
\label{sec:main.dist}

We want to define a distance between temporal graph embeddings satisfying the requirements stated above. Notice that the embedding vectors $\mathbf{x}$ are defined up to an orthogonal transformation applied to the rows of $X$. In fact, suppose $R \in \R^{d\times d}$ is an orthogonal matrix, and $\tilde{\mathbf{x}}_i = R\mathbf{x}_i$, then $\mathbf{x}_i^T\mathbf{x}_j = \mathbf{\tilde{x}}_i^T\mathbf{\tilde{x}}_j$ and the loss function of Eq.~\eqref{eq:EDRep} is $\L_P(\tilde{X}) = \L_P(X)$.
Hence, our distance must be invariant under orthogonal transformations. Moreover, as discussed above, the distance for unmatched graphs must also be invariant with respect to node permutations.

\medskip

Given these requirements, we now discuss separately our choices for the the matched and unmatched cases. In the \nameref{sec:empirical} section, we show the ability of our distances to discriminate between temporal graph classes that differ for topological or temporal properties.

\subsubsection*{Distance definition for matched graphs}

Let $X_1$ and $X_2 \in \R^{n \times d}$ be the embedding matrices of two matched temporal graphs $\G_1$ and $\G_2$. To satisfy the invariance requirements discussed above, rather than defining our distance directly in terms of the matrices $X_1$ and $X_2$, we use the auxiliary matrices $M_{X_1} = X_1X_1^T$ and $M_{X_2} = X_2X_2^T$ of size $n \times n$. 
These matrices are by construction invariant with respect to orthogonal transformations of the embedding space and encode the {global} structure of the corresponding temporal graphs, as the embeddings are generated using the complete temporal network information.
We define the distance $d_{\rm m}$ between matched temporal graphs as:
\begin{align}
 d_{\rm m}(\G_1, \G_2) := \Vert M_{X_1} - M_{X_2} \Vert_{\rm F} \,\, ,
\end{align}
where $\Vert\cdot\Vert_{\rm F}$ is the Frobenius norm. 
Each matrix $M$ encodes the similarity in the embedding space between all pairs of nodes of the corresponding temporal graph, and $d_{\rm m}$ quantifies thus the difference between these similarity patterns. In the \nameref{sec:supp} section, we show that we do not need to actually compute the large ($n \times n$) $M_{X_1}$ and $M_{X_2}$ matrices, but that the distance can be computed using the generally smaller $d\times d$ matrices as follows:
\begin{align}
    \label{eq:dm}
 d_{\rm m}(\G_1, \G_2) = \sqrt{\Vert X_1^TX_1\Vert_{\rm F}^2 + \Vert X_2^TX_2\Vert_{\rm F}^2 - 2\Vert X_1^TX_2\Vert_{\rm F}^2} \,\, ,
\end{align}
We can easily verify that $d_{\rm m}$ is a distance, as it satisfies all the requirements mentioned above and is generally not invariant under the permutation of node indices. 
Notice that, even if $d_{\rm m}$ can be computed for any pair of graphs with the same size, this distance is only meaningful for matched graphs (i.e., with known matching). In Eq.~\eqref{eq:dm} indeed, the term $X_1^TX_2$ implies that $X_1$ and $X_2$ have the same size and that their rows are ordered according to the same node indices.

\subsubsection*{Distance definition for unmatched graphs}

In the case of \emph{unmatched} graphs, there is no correspondence between nodes, and the distance, as we discussed, should also be invariant with respect to node permutations.
We therefore define the distance between unmatched graphs using auxiliary vectors. Specifically, we denote with $\bm{\lambda}(A)$ the vector of ordered eigenvalues of a matrix $A$ and define our distance between two \emph{unmatched} temporal graphs as
\begin{align}
    \label{eq:du}
 d_{\rm u}(\G_1, \G_2) = \left\Vert \bm{\lambda}\left(\frac{X_1^TX_1}{n_1}\right) - \bm{\lambda}\left(\frac{X_2^TX_2}{n_2}\right)\right\Vert_2 \,\, , 
\end{align}
where $X_1$ and $X_2$ are, as before, the embeddings matrices of the two graphs, with sizes $n_1 = |\V_1|$ and $n_2 = |\V_2|$. The normalization of the covariance matrices $X_1^TX_1$ and $X_2^TX_2$ is chosen so that their eigenvalues are independent from the graph size.

In the \nameref{sec:supp} section we show that the covariance matrices $X_1^TX_1$ and $X_2^TX_2$ (of size $d \times d$) are indeed an appropriate argument of the distance function $d_{\rm u}$, since they are independent of the ordering of the nodes and any other orthogonal transformation applied to the columns of $X_1$ and $X_2$.
However, a direct comparison of the entries of $X_1^TX_1$ and $X_2^TX_2$ would require a one-to-one correspondence between the embedding dimensions, i.e., the columns of $X_1$ and $X_2$. This mapping is unavailable because the \ptv algorithm, like most vector-space embedding techniques, makes no guarantees on the correspondence between the dimensions of the embedding vectors related to different graphs. 
To tackle this issue, we thus compare the spectra, which are invariant with respect to orthogonal transformations of the rows of the embedding matrices.
This choice takes inspiration from spectral distance definitions for static graphs, such as Ref.~\cite{torres2018graph}, in which graphs are compared by computing the distance between the ordered sets of the eigenvalues of their matrix representations (e.g., adjacency, Laplacian, non-backtracking, etc.)

\subsection*{Computational complexity}
As discussed in the \nameref{sec:supp}, the computational complexity of the embedding step is $\mathcal{O}\left(nd^2 + d\cdot{\rm min}(n^2, E)\right)$, where $E = \sum_t |\E_t|$ is the total number of temporal edges, $n$ is the number of graph nodes and $d$ is the embedding dimension. On top of this, the distance $d_{\rm m}$ is then obtained with $\O(nd^2 + d^2)$ additional operations, while $d_{\rm u}$ requires $\O(nd^2 + d^3)$ additional operations.

\subsection*{Evaluation with empirical data}
\label{sec:empirical}
We now evaluate the distance definitions we introduced, particularly with respect to their sensitivity to important topological and temporal properties of empirical temporal graphs. We carry out three kinds of evaluation tests 
using both synthetic and empirical temporal graph data.
We use, in particular, empirical data shared by the \texttt{SocioPatterns} collaboration (\href{http://www.sociopatterns.org/}{sociopatterns.org}) describing time-resolved, close-range proximity interactions of humans and animals in a variety of real-world environments
(see Table~\ref{tab:SP} of the \nameref{sec:supp} section). These data are known to exhibit rich multi-scale network structures, complex temporal activity patterns, correlations between these two, and other complex network features \cite{isella2011s}. Hence, they provide a natural benchmark to assess the sensitivity of our distance definitions to several specific properties of real-world temporal graphs. It is important to note that these datasets are by no means an exhaustive sample of temporal networks. However, our goal here is to show that the distances we introduce are sensitive to complex properties of temporal networks that we know are present in these particular datasets. Therefore, the evaluations we outline below do not depend on testing our distances on a comprehensive set of temporal networks but rather rely on making tests using data that are well understood, have been and are widely used by the scientific community and are known to exhibit many important properties of temporal graph data.

\begin{enumerate}
    \item \nameref{sec:main.test_synth}. We consider topologically different synthetic temporal graphs of different sizes and we cluster them using our temporal graph distance $d_{\rm u}$, checking how well the clusters we find match the known classes of the generative method we used, irrespective of graph size. We then carry out the same test using empirical graphs on human proximity in school settings.
    \item \nameref{sec:node_relabel}. We consider a synthetic temporal graph, select a fraction of its nodes, shuffle their indices and compute the distance $d_{\rm m}$ between the original and the re-labeled graphs. We investigate the behavior of the distance $d_{\rm m}$ as a function of the fraction of re-labeled nodes.

    \item \nameref{sec:discriminate_rdn}. We use the empirical data to elucidate which of the many properties of real temporal graphs are discriminated by the distances we introduced. We proceed as follows: given an empirical temporal graph, we apply to it a set of randomization operations, where each operation is designed to preserve specific properties of the original data (e.g., the activity time series of each node). We then attempt to discriminate the randomized temporal graphs from one another using our distance definitions and quantify the performance of such discrimination tasks. We test both $d_{\rm m}$ and $d_{\rm u}$; however, by design, this evaluation strategy compares temporal graphs with the same number of nodes.

\end{enumerate}

\subsubsection*{Discriminating between classes of temporal graphs}
\label{sec:main.test_synth}

To test the ability of our model to distinguish topological network features, we proceed as follows. We consider four random generative models for static graphs and generate $250$ instances from each model with size uniformly distributed between $n = 200$ and $n = 1800$ and constant average degree. The models we use are:
i) the Erd\H{o}s-Renyi (ER) model \cite{erdHos1960evolution};
ii) the stochastic block model (SBM), capable of generating graphs with a known community structure \cite{karrer2011stochastic};
iii) the configuration model (CM), yielding graphs with an arbitrary degree distribution \cite{bollobas1980probabilistic};
iv) the geometric model (GM), which generates edges based on proximity in a latent space \cite{dall2002random}. Unlike the first three, the latter model yields graphs with a high clustering coefficient. Details on these generative models are provided in the \nameref{sec:supp} section. 
We then turn these graphs into temporal graphs by assigning an activity time series to each edge of the static graph, {\ie~a set of timestamps at which that edge is active}. To do this, we copy the activity time series of a randomly chosen edge of an empirical temporal graph, namely the \emph{Conference} dataset, which has the largest number of nodes among the empirical datasets we use. Details on the properties of this dataset are given in Table~\ref{tab:SP}.
We then deploy our distance definition $d_{\rm u}$ to compute a pairwise distance matrix between all these temporal graphs and use this matrix as an input for an unsupervised clustering algorithm to label each graph. The unsupervised clustering algorithm only takes the distance matrix and the number of clusters $k$ (here $k=4$) as inputs. Here, we first perform a spectral embedding of the distance matrix between graph pairs using non-negative matrix factorization (NMF)~\cite{fevotte2011algorithms} extracting $k$ non-negative components and then applying the k-means clustering algorithm with $k$ classes~\cite{macqueen1967some}. We note that successively embedding and clustering is a standard approach, and using NMF is most suited here since the distance matrix is non-negative. Finally, we compare the original temporal graph labels (i.e., the classes of our generative model) with the partition labels yielded by unsupervised clustering based on our distances: we quantify the match between the true partition and the inferred one using the normalized mutual information (NMI), which yields a performance metric ranging between $0$ (random guess) and $1$ (perfect reconstruction).
\begin{figure*}[!t]
	\centering
	\includegraphics[width=\linewidth]{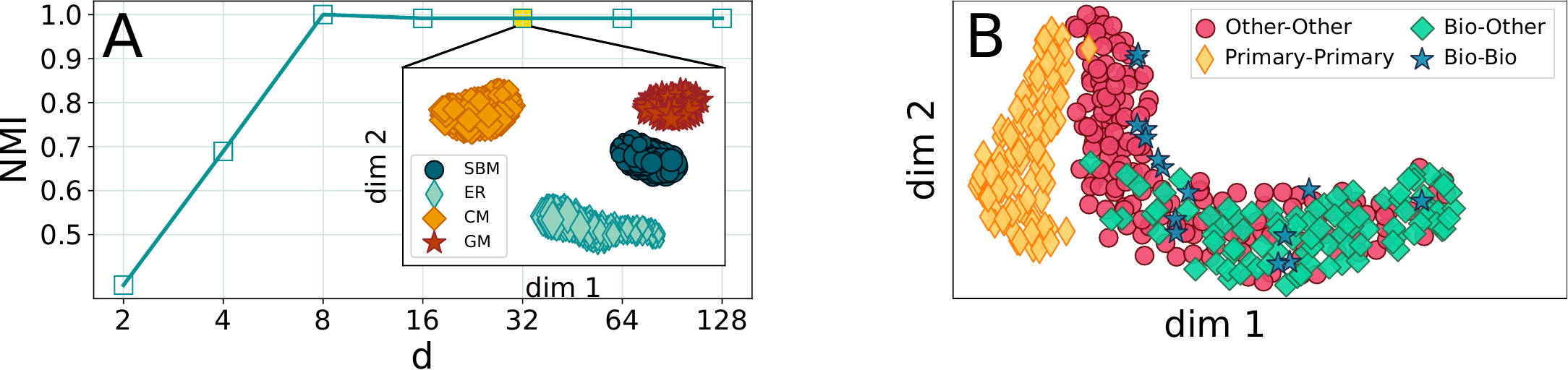}
	\caption{\textbf{Validation of the distances on graphs of varying size}.
		\emph{Panel A}: Accuracy of the distance-based clustering against the ground truth classes in terms of normalized mutual information (NMI), as a function of the embedding dimension $d$ used in the embedding step shown in Figure~\ref{fig:pipeline}.
		The clustering task consists in recognizing four classes of synthetic temporal graphs with an unsupervised algorithm based on the distance $d_{\rm u}$. The classes are obtained by generating a graph from either of four models (stochastic block model (SBM), configuration model (CM), Erd\H{o}s-Renyi (ER), and geometric model (GM)) with constant degree equal to $4.8$, and the temporal component is obtained by sampling the edge activity of an empirical graph, as detailed in the main text.
		\emph{Inset of panel A:} Scatter plot of \texttt{UMAP} dimensionality reduction in two dimensions of the vector $\bm{\lambda}$ appearing in the definition of $d_{\rm u}$ given in Eq.~\eqref{eq:du}, with $d = 32$. Each point refers to a temporal graph; the color and marker style refer to the generative model of its static component, while the marker size is proportional to the number of nodes.
		\emph{Panel B}: $2$-dimensional \texttt{UMAP} embedding of $\bm{\lambda}$ for the temporal graph obtained selecting two classes and a day of interaction for all possible $(c_1, c_2, \mbox{Day})$ triplets in the \texttt{SocioPatterns} datasets describing temporal graphs of human proximity in schools. Each point refers to a triplet and the color and marker style are assigned according to the class labels: the primary school classes form a group on their own and the other three groups (\emph{Other-Other}, \emph{Bio-Other}, \emph{Bio-Bio}) belong to the high school datasets, where \emph{Bio} are the biology classes, and \emph{Other} the remaining ones. In all cases, the temporal networks are aggregated over a scale of $t_{\rm res} = 10$ minutes.
	}
	\label{fig:clustering}
\end{figure*}

\medskip

Figure~\ref{fig:clustering}A shows the normalized mutual information (NMI) between the known temporal graph classes (associated with the four generative models) and the inferred cluster labels, as a function of the embedding dimension $d$. 
The discrimination performance is low for small values of $d$, but for $d \gtrsim 8$ the accuracy is high and becomes insensitive to the specific value of $d$.
The inset in Figure \ref{fig:clustering}A shows the \texttt{UMAP} \cite{sainburg2021parametric} $2$-dimensional embedding of the vectors $\bm{\lambda} \in \R^{32}$ used in the distance $d_{\rm u}$. Each symbol corresponds to one temporal graph, the color and marker shape indicate its generative model and the marker size is proportional to the number of graph nodes. We observe that the graph size does not appear to affect distances systematically and that the $\bm{\lambda}$ vectors used by the distance $d_{\rm u}$ allow us to discriminate between all four generative models. Indeed, we verified that the distances $d_{\rm u}$ between graph pairs and the Euclidean distances between their respective \texttt{UMAP} embeddings are strongly correlated, with a highly significant Spearman correlation coefficient of $\sim0.92$.

\medskip

We now carry out a test using the distance $d_{\rm u}$ on empirical temporal graphs. We look at the challenging setting in which the graph pairs considered have comparable sizes and time spans and are collected in similar contexts. To do so, we use the \texttt{SocioPatterns} datasets collected in schools \emph{Primary school, High school 1, High school 2, High school 3}, described in Table~\ref{tab:SP}, representing time-resolved, close-range proximity interactions of individuals grouped in multiple school classes, over several days. We extract sub-temporal graphs from these datasets and compare them with one another using the distance $d_{\rm u}$. We associate the school classes with a label: \emph{Primary} for all primary school classes; \emph{Bio} for the high school classes on biology; \emph{Other} for the remaining classes.

We consider a pair of classes $(c_1, c_2)$ from the same dataset, and we build a sub-graph restricted to all the interactions on a given day ``$\mbox{Day}$'' between the students of the classes $c_1$ and $c_2$. We finally compute the $2$-dimensional \texttt{UMAP} embedding from the vector $\bm{\lambda}$ appearing in Eq.~\eqref{eq:du}. Figure~\ref{fig:clustering}B shows the result of this procedure for all possible triplets $(c_1, c_2, \mbox{Day})$. Class labels are indicated by the markers' color and style. We observe a very high correlation between the distances $d_{\rm u}$ of graph pairs and the Euclidean distances between their \texttt{UMAP} embedding vectors, with a Spearman correlation of $~0.88$. That is, our distance definition can tell apart the networks extracted from the primary school dataset from those coming from the high school dataset. Moreover, a weaker but still visible separation exists in the embeddings of (\emph{Bio-Other}) with respect to the other triplets. This is likely because biology classes were often held in school laboratories rather than in classrooms, leading to different interaction patterns. Our distance can discriminate between temporal networks that are, in principle, similar (school classes over one day) with no major structural or temporal differences.
\begin{figure*}[!t]
	\centering
	\includegraphics[width=\linewidth]{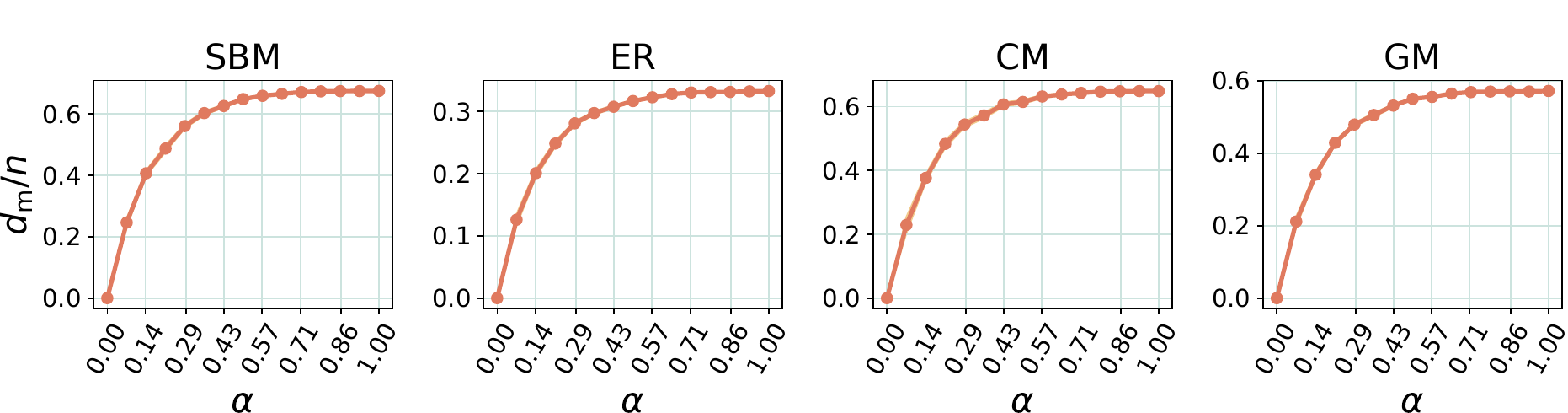}
	\caption{\textbf{Detection of partial node relabeling}.
		Normalized matched distance $d_{\rm m}/n$ between a temporal graph and itself, upon partial node re-labeling, as a function of the fraction $\alpha$ of re-labeled nodes. The plots refer to the four generative models used in Figure~\ref{fig:clustering}A: stochastic block model (SBM), Erd\H{o}s-Renyi (ER), configuration model (CM), and geometric model (GM). with average degree equal to $4.8$. For each graph, the temporal component is obtained by sampling the edge activity of an empirical graph, as detailed in the main text. The barely visible shadow line is the standard deviation of the distance across $25$ randomizations of the partial re-labeling.
	}
	\label{fig:relabeling}
\end{figure*}

\subsubsection*{Detecting partial node relabeling}
\label{sec:node_relabel}
One required property for matched distance is to distinguish between graphs that differ only upon a node re-labeling. To evaluate the ability of our distance to accomplish this task, we work with the synthetic temporal graphs defined in section~\nameref{sec:main.test_synth}. For each random graph model, we
i) generate an instance of the random graph with $n = 1000$ nodes;
ii) randomly select a fraction $\alpha$ of nodes and shuffle their labels;
iii) compute the distance between the original graph and the partially re-labeled one.
This is repeated for $25$ different realizations of the partial re-labeling. Figure~\ref{fig:relabeling} shows the distance $d_{\rm m}$ (rescaled by the graph size) as a function of $\alpha$. The plots confirm that the distance depends on partial node re-labeling for all the considered generative models, with a positive distance as soon as $\alpha >0$, and increases with the fraction $\alpha$ of relabeled nodes. Notably, the plot suggests a high discriminative power even for small $\alpha$ values, as the distance increases rapidly with $\alpha$.

\subsubsection*{Discriminating between temporal graph randomizations}
\label{sec:discriminate_rdn}
Let us now consider $9$ empirical \texttt{SocioPatterns} datasets, whose basic properties are summarized in Table~\ref{tab:SP} of the \nameref{sec:supp} section. Following Ref.~\cite{gauvin2018randomized}, we select $6$ types of randomization operations $R_i$ ($i=1, \ldots, 6$), described in detail in the \nameref{sec:supp}  section. In all cases, we preserve the number of nodes and the number of snapshots of the original temporal graph. These six randomizations are chosen because they allow us to inspect relevant temporal graph properties, including the interaction duration distribution, the node activation time series, the strength distribution in the aggregated graph, or the presence of a community structure. We can thus investigate graph properties both at the topological and temporal levels.
\begin{figure*}[!t]
    \centering
    \includegraphics[width=\linewidth]{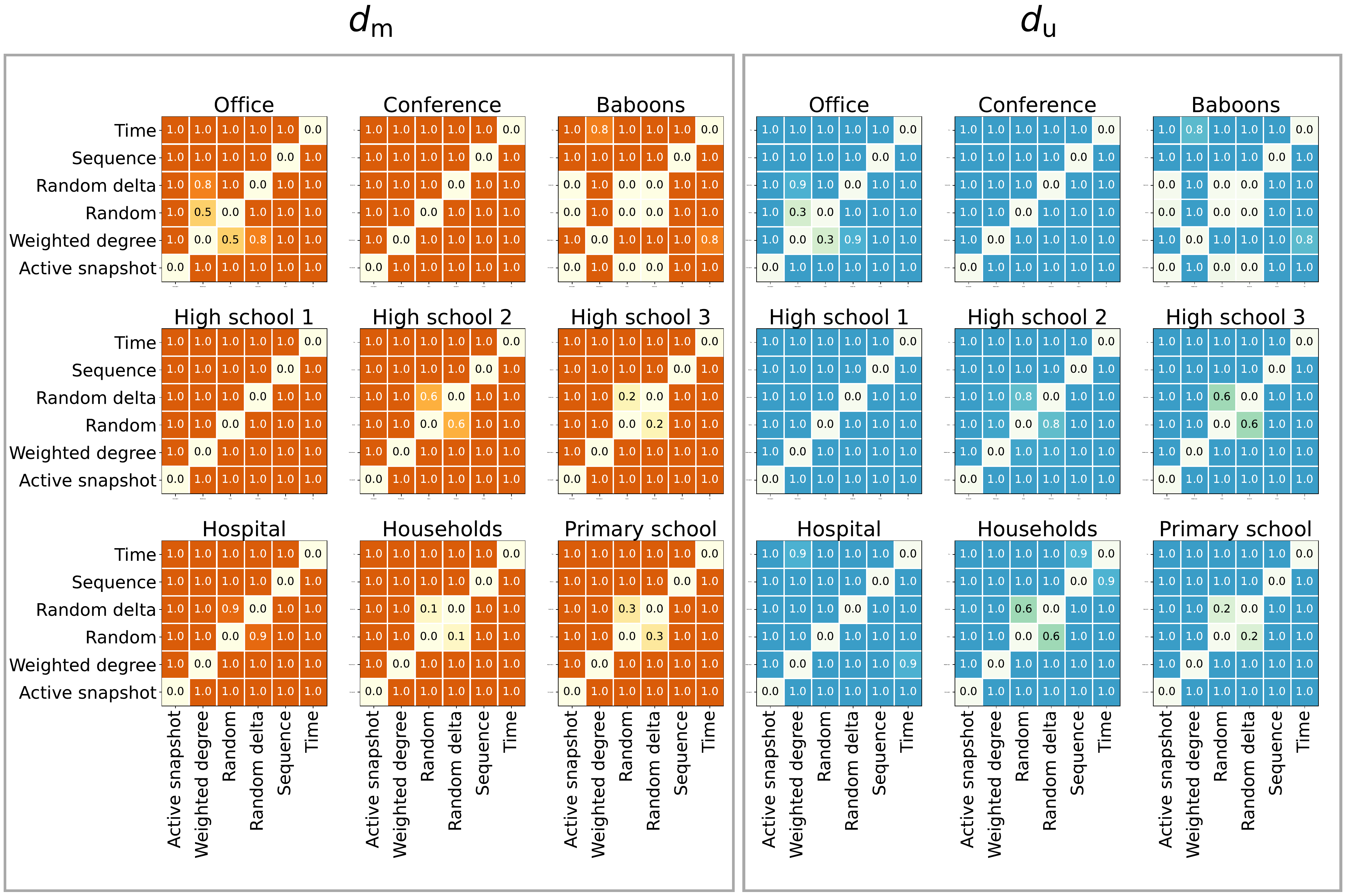}
    \caption{\textbf{Distance-based graph clustering for ensembles of temporal graphs {generated} according to different randomization techniques.} 
 \emph{Left panel}: results for the matched distance $d_{\rm m}$ of Eq.~\eqref{eq:dm}. \emph{Right panel}: results for the unmatched distance $d_{\rm u}$ of Eq.~\eqref{eq:du}. Within each panel, each matrix corresponds to one of the $9$ \texttt{SocioPatterns} temporal graphs described in Table~\ref{tab:SP}.
 The rows and columns of each matrix correspond to the same set of $6$ randomization techniques we used, described in the \nameref{sec:supp}  sections. Each input graph is represented as a sequence of temporal edges $(i,j,t)$ as per Definition~\ref{def:dynamic}. The randomizations act on the temporal edges as follows.
        \emph{Random}: preserves the number of temporal edges and randomizes the node and time indices;
        \emph{Random delta}: preserves the number of temporal edges and interaction duration distribution;
        \emph{Active snapshot}: preserves the number of edges at each time-step and the times at which each node is \emph{active}, i.e., at which it has at least one neighbor.
        \emph{Time}: preserves the aggregated weighted graph structure, i.e., the number of times each edge is active in the temporal graph;
        \emph{Sequence}; preserves each snapshot's adjacency matrix and randomizes the order in which they appear;
        \emph{Weighted degree}: preserves the total number of temporal edges involving each node.
 For each pair of randomizations, we infer the randomization method of each temporal graph via an unsupervised distance-based clustering algorithm, and we compare the inferred randomization method with the known true one. Each matrix entry reports (value and color coding) the accuracy of the inferred labels (randomization methods), quantified as the normalized mutual information (NMI) between the inferred and true labels.
 }
    \label{fig:shuffle}
\end{figure*}

To quantify the discriminative performance of our temporal graph distance, we proceed as follows, separately for each empirical temporal graph: given the empirical temporal graph and a pair of randomization operations ($R_i$, $R_j$), we generate a set of $250$ realizations using $R_i$, labeled as $0$, and a set of $250$ realizations using $R_j$, labeled as $1$, for a total of $500$ temporal graphs.
We then compute the matrix of distances between these $500$ graphs and use it to cluster them in $2$ clusters, following the same procedure used for Fig.~\ref{fig:clustering}A.
Figure~\ref{fig:shuffle} reports the NMI values between the real labels and the unsupervised algorithm result
for each pair of randomizations ($R_i$, $R_j$),
and for both the matched distance $d_{\rm m}$ (left panel) and the unmatched distance $d_{\rm u}$ (right panel). Within each panel, each matrix refers to one empirical temporal graph, and its entries are the NMI values described above.
{A high NMI entry indicates that our distance discriminates well between the temporal graphs generated according to the two randomization procedures in the corresponding row and column of the matrix entry, i.e., that the distance is sensitive to the properties preserved by either $R_i$ or $R_j$.}
Low NMI values may either correspond to an inability of the distance to capture those properties or simply reflect that the empirical temporal graph is statistically random as far as those properties are concerned, {as observed in particular for several pairs of randomizations in the \emph{Baboons} dataset}.
Thus, it is enough to find one empirical dataset for which the
distance can distinguish between a pair of randomizations to conclude that the distance is indeed sensitive to the corresponding graph properties.

\section*{Discussion}
We introduced a novel definition of distance between pairs of temporal graphs. This definition entails two steps. 
First, the temporal graphs are embedded in Euclidean space, and then a distance is defined in embedding space. For the first step, we use an embedding based on time-respecting random walks over the temporal graph. Such walks are known to depend on and to encode important structural and temporal properties of time-varying graphs \cite{pan2011path,holme2012temporal,sato2019dyane,zhan2021measuring, saramaki2015exploring}. 
For the second step, we proposed two possibilities for the distance definition: if a mapping is known between the nodes of the graphs to be compared, we consider a distance definition that leverages such mapping; in the more general case, when such a mapping is unavailable, we put forward a definition that makes it possible to compare graphs with arbitrarily different sizes (numbers of nodes).
In both cases, since the size of the embedding matrix we use does not depend on the graph's temporal span, it is possible to embed temporal graphs with different durations in the same embedding space and thus compute a distance between them.
The evaluation of our approach on both synthetic and empirical data shows that the proposed distance is sensitive to structural differences (e.g., degree distribution, clustering coefficient, or presence of communities) as well as to temporal differences (e.g., burstiness or node activity patterns) of the temporal graphs being compared.

\medskip

The methodology we developed is very general and customizable, as several alternative choices are possible for both steps of our approach, namely the chosen embedding method and distance. The specific choices we study here are motivated by two considerations:
i) a theoretical one, given the importance of 
time-respecting random walks in encoding information on temporal graphs, and thus in generating distributed representations of temporal graphs~\cite{pan2011path,holme2012temporal,sato2019dyane, saramaki2015exploring};
ii) a numerical one, which led us to choose the \texttt{EDRep} algorithm of Ref.~\cite{dall2023efficient} because of its efficiency and scalability.
Our evaluation results support these choices a posteriori.

Some limitations of our approach are worth mentioning.
First, the part of our evaluation based on empirical data focused on a limited set of temporal graphs, namely temporal networks describing the close-range proximity of humans. However, these temporal graphs are  known to feature a broad variety of representative structural and temporal features, such as time-varying community structures, burstiness of edge activity, fat-tailed distributions of interaction durations, and more \cite{cattuto2010dynamics,barrat2013temporal}. 
Second, we only considered the cases of fully matched graphs (a bijective relation between nodes) or no known matching. An interesting intermediate situation would be partially matched graphs, in which only a subset of nodes are matched across the two graphs. Tackling this challenging case would yield an interesting extension of our work. Finally, our unmatched distance allows us to compare graphs with sizes (number of nodes) that potentially differ even by orders of magnitude. Even if our definition can deal with such extreme cases, we believe that this kind of comparison calls for a more profound question on what it means to compare entities that differ so much. This case might require tailored definitions of distance that leverage domain-specific knowledge.

\medskip

Defining a distance between temporal graphs opens the door to a wealth of potential applications, of which our evaluation only offers possible examples. 
On the one hand, such a distance can support data analysis of temporal graph data by extending the procedure of Figure~\ref{fig:shuffle} 
to a richer set of randomizations~\cite{gauvin2014detecting}. Indeed, as discussed, the inability to distinguish pairs of randomizations of a given temporal graph can be regarded as an invariance property of the graph itself.
On the other hand, computing a distance between graphs can provide a crucial tool for the validation of generative models that is often limited to comparing a set of statistical properties~\cite{bail2023flow}, while our distance definition enables us to carry out a comparison at the global level.
In addition, our distance could underpin the very process of generating synthetic temporal graphs by enabling approaches such as \emph{GANS}~\cite{goodfellow2014generative} that depend on a global distance and have proved extremely powerful in generating realistic data~\cite{faez2021deep, alqahtani2021applications}. In particular, generating synthetic human proximity data with realistic topological and temporal properties could be used to better simulate infectious disease spread and other phenomena of interest for public health research~\cite{Machens2013,gemmetto2014mitigation,mistry2021inferring, cencetti2021digital, crectu2022interaction,COLOSI2022977}. 
Finally, we know that empirical (temporal) graphs are hard to anonymize~\cite{backstrom2007wherefore, romanini2021privacy, crectu2022interaction}. To minimize the risk of node re-identification, performing some perturbation operations on the graph might be necessary before making it publicly available. However, such operations risk destroying essential patterns and information of empirical data. The distance we introduced could help tackle the trade-offs between re-identification risk and information loss by better quantifying the latter.

\section*{Methods}
\label{sec:supp}

\subsection*{The \ptv algorithm}
\label{sec:supp.p2v}

The \ptv algorithm~\cite{dall2023efficient} {was recently proposed to efficiently generate embeddings given a probability matrix encoding the affinity between the embedded items. This is done by optimizing the cost function of Eq.~\eqref{eq:EDRep} under the constraint $\Vert \mathbf{x}_i \Vert = 1$ and obtaining a low-dimensional representation of matrix $P$. A known problem of this type of cost function is the computational complexity because the normalization constants $Z_i = \sum_{k\in\V} e^{\mathbf{x}_i^T\mathbf{x}_k}$ require $\O(n^2)$ to be calculated.} However, Ref.~\cite{dall2023efficient} describes an efficient way to estimate all the $Z_i$ values in $\O(n)$. This is accomplished by first subdividing the nodes into $q$ groups based on the embedding matrix $X$. Here $q$ is a parameter of the algorithm, and larger $q$ values generally lead to a higher accuracy, but very good results are already obtained for $q=1$. Second, for each $a = 1,\dots,q$, one computes the mean $\bm{\mu}_a$ and the covariance matrix $\Omega_a$ of the set of embedding vectors of the nodes in group $a$. 
Denoting the number of nodes in group $a$ by $\pi_a$, one obtains the estimation of $Z_i$ as
\begin{align}
	Z_i \approx \sum_{a = 1}^{q} \pi_a
	~ {\rm exp}\left\{\mathbf{x}_i^T\bm{\mu}_a + \frac{1}{2}\mathbf{x}_i^T\Omega_a\mathbf{x}_i\right\}\,\, .
\end{align}
Coming to the algorithm's computational complexity, given that our choice of $P$ is the sum of matrix products, the algorithm can be implemented in two forms: one in which $P$ is never formally computed and the sequence $\{\hat{L}_t\}_{t = 1,\dots,T}$ is fed to the algorithm; and one in which $P$ is explicitly computed. Letting $X \in \R^{n\times d}$ be the matrix storing the embedding vectors in its rows, the computational complexity is determined by the matrix product $PX$ needed to obtain the gradient of $\L_P$ defined in Eq.~\eqref{eq:EDRep}. In the former case, this is computed in $\O(nd^2 + dE)$ operations, where $E = \sum_{t=1}^T |\E_t|$ is the number of temporal edges. In the second case, instead, the complexity is given by the number of non-zero entries of $P$, which cannot exceed $n^2$. The former implementation is particularly convenient when $n$ is large and the $\hat{L}_t$ matrices are very sparse. This happens because $P$ may be dense, even if the snapshots are sparse, and for large values of $n$, representing the matrix $P$ in computer memory might be challenging. Conversely, for small graphs, $n^2$ may be smaller than the number of temporal edges, making the latter implementation more convenient.

\subsection*{Matched distance definition}
\label{sec:methods.dm}

From Eq.~\eqref{eq:dm}, we have
\begin{align}
	&d^2_{\rm m}(X, Y) = \sum_{i,j\in\V}\left[(M_X)_{ij} - (M_{Y})_{ij}\right]^2\nonumber\\
	&= \sum_{i,j\in\V} 
	\left[ (M_X)_{ij}(M_X)_{ij} + (M_Y)_{ij}(M_Y)_{ij} - 2(M_X)_{ij}(M_Y)_{ij} \right]\nonumber\\
	&\overset{(a)}{=} \sum_{i,j\in\V}
	\left[ (M_X)_{ij}(M_X)_{ji} + (M_Y)_{ij}(M_Y)_{ji} - 2(M_X)_{ij}(M_Y)_{ji} \right] \nonumber\\
	&= \sum_{i\in\V} \left[ (M_XM_X^T)_{ii} + (M_YM_Y^T)_{ii} - 2(M_XM_Y^T)_{ii} \right] \nonumber\\
	&= {\rm tr}(M_XM_X^T) + {\rm tr}(M_YM_Y^T) - 2{\rm tr}(M_XM_Y^T)\nonumber\\
	&\overset{(b)}{=} {\rm tr}(XX^TXX^T) + {\rm tr}(YY^TYY^T) - 2{\rm tr}(XX^TYY^T)\nonumber\\
	&\overset{(c)}{=} {\rm tr}(X^TXX^TX) + {\rm tr}(Y^TYY^TY) - 2{\rm tr}(Y^TXX^TY)\nonumber\\
	&\overset{(d)}{=} \Vert X^TX \Vert_{\rm F}^2 + \Vert Y^TY\Vert_{\rm F}^2 - 2 \Vert X^TY \Vert_{\rm F}^2 \,\, ,
\end{align}
where in $(a)$ we exploited that $M_X$ and $M_Y$ are symmetric, in $(b)$ we used the definition of $M_X, M_Y$, in $(c)$ we used that property of the trace stating that ${\rm tr}(AB) = {\rm tr}(BA)$, and finally in $(d)$ we applied the definition of the Frobenius norm.

\subsection*{Properties of the unmatched distance}
In this section, we show that the unmatched distance is invariant with respect to node permutations, and, more generally, to any orthogonal transformation applied to the columns of the embedding matrix. We then show that the unmatched distance is also invariant with respect to any orthogonal transformation applied to the rows of the embedding matrix. This is necessary because the \texttt{EDRep} loss function is invariant under this type of transformation.

\medskip 

We consider a bijective node mapping $\pi: \V \to \V$. We let $Q\in\R^{n\times n}$ be the permutation matrix defined as $Q_{ij} = \delta_{j, \pi(i)}$. In the matrix multiplication, $Q$ swaps the entry $i$ with $\pi(i)$:
\begin{align}
\bar{X}_{ia} := \left(QX\right)_{ia} = \sum_{j\in\V} Q_{ij}X_{ja} = \sum_{j\in\V} \delta_{j,\pi(i)}X_{ja} = X_{\pi(i),a}\,\, .
\end{align}
The matrix $Q$ is orthogonal, in fact
\begin{align}
(QQ^T)_{ij} = \sum_{k\in\V} Q_{ik}Q_{jk} = \sum_{k\in\V} \delta_{k,\pi(i)}\delta_{k,\pi(j)} = \delta_{\pi(i),\pi(j)} = \delta_{ij}\,\, ,
\end{align}
where the last equality follows from $\pi$ being a bijective mapping. As a consequence, $QQ^T = Q^TQ = I_n$. The distance $d_{\rm u}$ depends on the embedding matrix $X$ only through $X^TX$. We now show that this expression is invariant under node permutation of the embedding matrix, or any orthogonal matrix $Q$. Indeed, let $\bar{X} = QX$, then
\begin{align}
\bar{X}^T\bar{X} = X^T\underbrace{Q^TQ}_{I_n}X = X^TX\,\, .
\end{align}
We now show that the distance $d_{\rm u}$ is also invariant under the orthogonal transformation applied to the embedding matrix rows. Let $R\in\R^{d\times d}$ be an orthogonal matrix and let $\tilde{X} = XR$. Letting $\lambda_i(M)$ be the $i$-th smallest eigenvalue of $M$, then, for all $i$, $\lambda_i(AB) = \lambda_i(BA)$ \cite[Theorem 1.3.22]{horn2012matrix}. It follows
\begin{align}
\lambda_i(\tilde{X}^T\tilde{X}) = \lambda_i(R^TX^TXR) =  \lambda_i(\underbrace{RR^T}_{I_d}X^TX) = \lambda_i(X^TX)\,\, ,
\end{align}
thus proving the claim.

\subsection*{Temporal graph randomizations}
\label{sec:supp.shuffle}

In Table~\ref{tab:SP} we summarize the properties of the temporal graphs we used to conduct our tests. Here we give a more detailed description of the randomization techniques we adopted, defined in terms of the quantities they preserve. According to the method, it may be more convenient to represent the temporal graph as a sequence of instantaneous interactions $(i,j,t)$, as a sequence of interactions with a duration $(i,j,t,\tau)$, or as a sequence of weighted adjacency matrices $W^{(t)}$~\cite{gauvin2018randomized}. Before randomization, time is discretized at the scale of $10$ minutes in each dataset, and the cumulative interaction time in each $ 10$-minute window is used as weight.

\begin{enumerate}
	\itemsep 0em
	\item \emph{Random}. Temporal edges are represented as $(i,j,t)$ and all three indices are randomized, avoiding self-edges ($i$ and $j$ randomly sampled with replacement between $1$ and $n$, and $t$ between $1$ and $T$). \\
	\textbf{Preserved quantities}: number of temporal edges.
	\item \emph{Random delta}. Temporal edges are represented as $(i,j,t,\tau)$. Once again, $i,j,t$ are randomized ($t$ is randomly sampled with replacement between $1$ and $T-\tau$), while $\tau$ is preserved.\\
	\textbf{Preserved quantities}: number of temporal edges and interaction duration distribution.
	\item \emph{Active snapshot}. At each time step $t$, the edges are randomly replaced between active nodes at $t$, \ie~that had at least one neighbor in the original snapshot.\\
	\textbf{Preserved quantities}: number of edges at each time-step and activity pattern of each node.
	\item \emph{Time}. Temporal edges are represented as $(i,j,t)$ and only the index $t$ is randomly sampled with replacement.\\
	\textbf{Preserved quantities}: aggregated graph structure.
	\item \emph{Sequence}. The graph is represented as a sequence of weighted adjacency matrices $W^{(t)}$ and the indices $t$ are shuffled.\\
	\textbf{Preserved quantities}: the structure of each snapshot.
	\item \emph{Weighted degree}. Temporal edges are represented as $(i,j,t)$ and all three indices are randomized as in \emph{Random} but with the constraint that each node appears in the same number of temporal edges as in the original network.\\
	\textbf{Preserved quantities}: nodes weighted degree.
\end{enumerate}

\begin{table}[!t]	
	\caption{{Summary properties of the \texttt{SocioPatterns} time-resolved proximity networks used here. \emph{Graph name} is used to identify the graphs; \emph{Description} provides concise information on the context where data were collected; $n$ is the number of graph nodes; \emph{Duration} is the temporal span of the dataset, and $T$ is the number of graph snapshots. The temporal resolution is $t_{\rm res} = 10$~min for all datasets.}}
	\centering
	\begin{tabular}{llrrr}
		Graph name & Description & $n$ & Duration & T \\
		\toprule
		\midrule
		\emph{Primary school}  &  Children of $10$ classes & $242$ &$2$ days & $194$\\
		\cite{stehle2011high, gemmetto2014mitigation}	& of a primary school& &&\\
		\midrule
		\emph{High school 1} &  Students of $3$ classes & $126$ & $4$ days & $453$\\
		\cite{fournet2014contact}& of a high school in $2011$&  &&\\
		\midrule
		\emph{High school 2} &  Students of $5$ classes & $180$ & $7$ days&$1215$\\
		\cite{fournet2014contact}& of a high school in $2012$& &&\\
		\midrule
		\emph{High school 3} &  Students of $9$ classes & $327$ & $5$ days&$605$\\ 
		\cite{mastrandrea2015contact}& of a high school in $2013$& &&\\
		\midrule
		\emph{Baboons} \cite{gelardi2020measuring} &  A group of baboons  & $13$ & $27$ days&$3986$\\ 
		\midrule
		\emph{Households} \cite{ozella2021using}  &  People of a village & $86$ & $26$ days & $1926$\\
		& in Malawi& & & \\
		\midrule
		\emph{Hospital} \cite{vanhems2013hospital}  &  Patients and health-care & $75$ & $5$ days & $579$\\
		& workers of a hospital & &\\
		\midrule	
		\emph{Conference} \cite{cattuto2010dynamics}  &  People at a medical & $405$ & $2$ days & $190$\\
		& conference  & & & \\
		\midrule	
		\emph{Office} \cite{genois2018can}  &  Office workers in an   & $92$ & $19$ days & $1646$\\
		& office building  & & & \\
		\midrule
		\bottomrule
	\end{tabular}
	\label{tab:SP}
\end{table}

\subsection*{Synthetic models}
\label{sec:app.synth}

We here provide a formal definition of the models used to generate the synthetic graphs under analysis. Even though we considered four models, three of these can be generated from the \emph{degree-corrected stochastic block model}~\cite{karrer2011stochastic} by changing its parameters.

\begin{definition}[Degree-corrected stochastic block model]
	Let $\V$ be a set of $n$ nodes and $\bm{\ell} \in [k]^n$ be a vector mapping each node to a class. Further let $C \in \R^{k\times k}$ be a positive symmetric matrix and $\bm{\theta} \in \R^{n}$ be a vector satisfying $\bm{\theta}^T\mathbf{1}_n = n$. The entries of the graph adjacency matrix $A \in\ [0,1]^{n\times n}$ are generated independently (up to symmetry) at random with probability
	\begin{equation}
		\P(A_{ij} = 1) = {\rm min}\left(1,\theta_i\theta_j\frac{C_{\ell_i, \ell_j}}{n}\right) \,\, .
	\end{equation}
\end{definition}

The vector $\bm{\ell}$ contains the labels and gives a community structure in the case in which $C_{a,a} > C_{a, b}$ for $b \neq a$, meaning that there is a higher probability that two nodes in the same community will get connected. The value $\theta_i$ is proportional to the expected degree of node $i$. For this reason, if one chooses $\bm{\theta} = \mathbf{1}_n$ and $\bm{\ell} = \mathbf{1}_n$, one gets the Erd\H{o}s-Renyi model, in which every node has the same expected degree, and there are no communities. The configuration model, instead, is obtained by letting $\bm{\ell} = \mathbf{1}_n$, but changing the value of $\bm{\theta}$ to create an arbitrary degree distribution that we choose to be a (properly rescaled) uniform distribution between $3$ and $10$ raised to the power $4$. Finally, the stochastic block model is obtained from a labeling vector different from $\mathbf{1}_n$ and letting $\bm{\theta} = \mathbf{1}_n$. We consider $k = 5$ communities of equal size with $C_{a,b} = 20 \delta_{ab} + (1-\delta_{ab})$, with $\delta$ the Kroeneker symbol.

\medskip 

Let us finally introduce the random geometric model.

\begin{definition}[Random geometric model]
	Let $\V$ be a set of $n$ nodes. For each $i\in\V$ let $\mathbf{x}_i \in \R^2$ be a random vector with norm $\Vert \mathbf{x}_i \Vert\leq 1$. The entries of the graph adjacency matrix $A \in [0,1]^{n\times n}$ are generated independently (up to symmetry) at random with probability
	\begin{align}
		\P(A_{ij} = 1) = e^{-\beta \Vert\mathbf{x}_i - \mathbf{x}_j\Vert} \,\, ,
	\end{align} 
	for some positive $\beta$.
\end{definition}

Note that even though the entries of $A$ are drawn at random, this model can generate graphs with a high clustering coefficient because the probability depends on the relative distance between fixed embedding vectors. In our simulations, we set $\beta = 20$. 

\section*{Data availability}

All data used in this article are publicly available at \href{http://www.sociopatterns.org/datasets/}{http://www.sociopatterns.org/datasets/} and at Ref.~\cite{dallamico2024GDynaDist}.

\section*{Code availability}

Commented \texttt{Python} code to reproduce our results is available at Ref.~\cite{dallamico2024GDynaDist}.


\section*{Acknowledgments}

LD and CC acknowledge support from the Lagrange Project of the ISI Foundation funded by CRT Foundation and from Fondation Botnar (EPFL COVID-19 Real Time Epidemiology I-DAIR Pathfinder).
AB acknowledges support from the Agence Nationale de la Recherche (ANR) project DATAREDUX (ANR-19-CE46-0008).

\section*{Authors contributions}

LD formalized the research problem, developed computer code, carried out the analysis, and wrote the first version of the manuscript. LD, AB and CC contributed to framing the research questions, discussing the results, and writing the manuscript.

\section*{Competing interests}

The authors do not declare any competing interests.

\end{document}